\documentclass[aps,prl,floatfix,twocolumn,superscriptaddress, showpacs]{revtex4}
\usepackage{color}
\usepackage{bm}
\usepackage{amssymb}
\usepackage{amstext}
\usepackage{graphics}
\usepackage{epsfig}
\usepackage{placeins}

\begin{document}

\title{Hidden SDW order and effective low-energy theory for FeAs
superconductors}
\author{Zheng-Yu Weng}
\affiliation{Center for Advanced Study, Tsinghua University, Beijing, 100084, China}
\date{\today }

\begin{abstract}
We propose a simple effective model to describe FeAs
superconductors. This model is based on the assumption of a
\emph{local} spin-density-wave (SDW) order, with its magnetization
direction allowed to fluctuate. It is shown that the long-range
order with momentum $\mathbf{Q}=(\pi ,\pi )$ is generally unstable
in competing with the kinetic energy of the charge carriers. A true
weak SDW order is formed in the undoped case with an additional
momentum shift $\mathbf{Q}_{s}=(\pi ,0)$ due to the peculiar Fermi
surface nesting. In the doped case, the fluctuating long-range order
driven by kinetic energy can naturally result in a d-wave
superconducting condensation. Such low-energy physics is protected
by the presence of the local SDW which sustains some kind of
\textquotedblleft Mott gaps\textquotedblright\ for the multiband
$d$-electrons near the Fermi energy.
\end{abstract}

\maketitle

\emph{Introduction. }The recent discovery of the iron-based superconductors%
\cite{yoichi,wen,chenxh,nlwang,ren} has stimulated a lot of interest
concerning underlying mechanism for superconductivity in this new
superconductor family. It has been established by various measurements\cite%
{zfang,cruz,McGuire} that there exists an SDW order in the undoped \textrm{%
LaOFeAs} compound below $T\sim 150$ $\mathrm{K}$, which quickly disappears
with the electrons doped into the system, where the superconducting phase
starts to set in with transition temperatures being raised beyond $50$ $%
\mathrm{K}$.\cite{ren,chenxh1} While the general phase diagram\cite%
{yoichi,wen,chenxh,nlwang,chenxh1} reminds us some interesting similarities
with the cuprate superconductors, the $d$-electrons on Fe seem much more
itinerant with multi-orbitals crossing the Fermi level as indicated by the
band structure calculations,\cite{lebegue,singh,kotlia,xu,cao,lu,kuroki}
compared to the isolated Cu$_{3d_{x^{2}-y^{2}}}$-O$_{2p_{x,y}}$ antibonding
orbitals in the cuprates. Many theoretical proposals are based on itinerant
approaches\cite{dai,zdwang,qi,lee} with the emphasis on the important role
played by various magnetic fluctuations, while some conjectures are also
made from the side of large-spin Mott insulators.\cite{li}

The LDA calculations\cite{cao,lu} have found an energetically \emph{robust}
SDW state at the antiferromagnetic (AF) momentum $\mathbf{Q}=(\pi ,\pi )$
with a large Fe moment $\sim 2.3\mu _{\mathrm{B}}$ per site, but
experimentally only a \emph{weak} SDW ordering with a different magnetic
momentum $\mathbf{Q}_{s}=$ $(\pi ,0)$, which further doubles the unit cell
of the former SDW state composed of two Fe per cell, has been identified\cite%
{cruz,McGuire} in the undoped case. The latter SDW (called stripe type
below) was predicted by the first principle band structure calculation\cite%
{zfang} due to the nesting Fermi surfaces of the hole and electron pockets,
which is much more \textquotedblleft fragile\textquotedblright\ and easily
destroyed as the doped electrons fill up the small hole pockets at small
doping.

In this paper, we will make a very simple proposal by assuming that the SDW
state with an AF momentum $\mathbf{Q}$ remains strong \emph{locally} in both
the undoped and small doped regime. The corresponding profile of the
electron density of states is schematically illustrated in Fig. 1. Based on
this minimal model, we can show that the long-range part of this SDW is
actually generically unstable, by coupling to the charge carriers near the
Fermi level. Such instability can result in a weak stripe-type SDW order in
the undoped case where the nesting Fermi surfaces are present, and naturally
a superconducting condensation when the former is destroyed at small doping.
In this whole regime, however, the local order of such an SDW remains robust
to \textquotedblleft protect\textquotedblright\ the low-energy physics,
which resembles the role of the \textquotedblleft Mott
gap\textquotedblright\ in the cuprate superconductors.\cite{anderson}

\begin{figure}[tbp]
\centerline{
    \includegraphics[height=1.8in,width=3.2in]{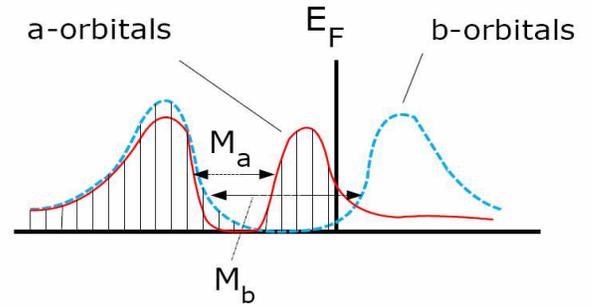}
    }
\caption{(color online) Schematic profile of the density of states near the
Fermi energy in the presence of the local SDW mean-fields $M_{a}$ and $M_{b}$%
, which decide the \textquotedblleft charge gaps\textquotedblright\ for the $%
a$-orbital and $b$-orbital bands according to $H_{0}$ [(\protect\ref{H0})].
The lower half of each band (the so-called $\protect\alpha $-band, see text)
is assumed below the Fermi energy $E_{F}$ and is filled, while the upper
half (the $\protect\beta $-band) is only partially filled. Here $a$-bands
refer to the bands passing the Fermi level, which are more d$_{xy}$, d$_{xz}$%
, d$_{yz}$ like, and $b$-bands are more of d$_{x^{2}-y^{2}}$ character
according to the LDA result.\protect\cite{cao,lu}}
\label{fig1}
\end{figure}

\emph{Minimal} \emph{Model. }In our model Hamiltonian $H_{\mathrm{eff}}=H_{%
\mathrm{band}}+H_{\mathrm{I}}$, the first term is a tight-binding model
\begin{equation}
H_{\mathrm{band}}=-\sum_{ij,a,b,\alpha }t_{ij}^{a,b}c_{ia\sigma }^{\dagger
}c_{jb\sigma }  \label{hband}
\end{equation}%
which describes the electron effective hoppings between the d-orbitals of
the Fe ions on square lattice (Fig. 2(a)) at the nearest neighboring (NN)
and next nearest neighboring (NNN) sites, including intra- and inter-orbital
hoppings with the superscripts $a,b$ specifying the orbitals. There have
been several proposals\cite{cao,kuroki,qi,lee} for $H_{\mathrm{band}}$ based
on the LDA calculations in order to capture the relevant bands near the
Fermi energy.

\begin{figure}[tbp]
\centerline{
    \includegraphics[height=1.8in,width=3.2in]{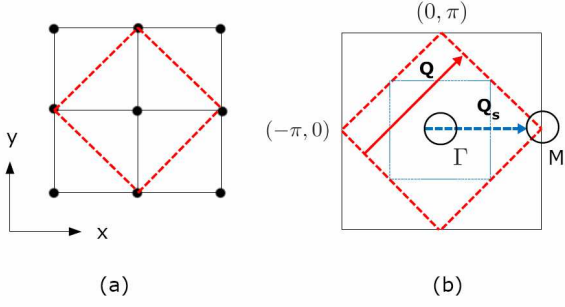}
    }
\caption{(color online) (a) The iron atoms form a square lattice with the
dashed diamond denoting the unit cell of the SDW order state with magnetic
momentum $\mathbf{Q}$. (b) The magnetic Brillouin zone (BZ) is illustrated
by the dashed diamond shape. Two circles at the $\Gamma $ and $M$ points are
hole and electron pockets connected by a momentum $\mathbf{Q}_{s}$. The
smaller square is a reduced BZ corresponding to an enlarged unit cell
indicated by the largest square in (a). }
\label{fig2}
\end{figure}

The second term $H_{\mathrm{I}}$ reflects the effective influence of the
Coulomb interaction on the $d$-electrons, which includes the on-site and NN
repulsions as well as the Hund's rule ferromagnetic coupling, and has the
following \textquotedblleft mean-field\textquotedblright\ look

\begin{equation}
H_{\mathrm{I}}=-\sum_{i,a}\mathbf{M}_{ia}\cdot \mathbf{S}_{ia}  \label{Hi}
\end{equation}%
where $\mathbf{M}_{ia}$ denotes an effective SDW mean-field felt by the
electron spin $\mathbf{S}_{ia}$ of the orbital $a$ at site $i.$ In an SDW
ordered state, one would have $\mathbf{M}_{ia}\propto \sum_{b(\neq
a)}J_{H}^{a,b}\left\langle \mathbf{S}_{ib}\right\rangle +U_{a}\left\langle
\mathbf{S}_{ia}\right\rangle $, where $J_{H}^{a,b}$ is the Hund's rule
coupling constant and $U_{a}$ is the on-site repulsion. We shall make the
following ansatz%
\begin{equation}
\mathbf{M}_{ia}=M_{a}(-1)^{i}\mathbf{n}_{i}  \label{M}
\end{equation}%
where a \emph{single} unit vector $(-1)^{i}\mathbf{n}_{i}$ will describe the
true polarization direction of $\mathbf{M}_{ia}$. Here the staggered factor $%
(-1)^{i}$ is introduced such that $\mathbf{n}_{i}=\mathbf{\hat{n}}$ will
correspond to a true AF order, but in a general case\ $\mathbf{n}_{i}$ will
not be fixed around a particular direction as only the relative change $%
\Delta _{\alpha }\mathbf{n}_{i}\equiv \mathbf{n}_{i+\hat{\alpha}}-\mathbf{n}%
_{i}$ ($\hat{\alpha}=\hat{x},\hat{y}$) will enter the Hamiltonian as shown
below.

The local SDW field $M_{a}$ will be assumed large according to the LDA
calculation,\cite{cao,lu} with the magnetization at different orbitals
tightly aligned together by the Hund's coupling. Under this assumption, the
long-wavelength, low-energy fluctuations of $\mathbf{n}_{i}$ may be treated
as an independent degree of freedom. $\mathbf{n}_{i}$ will be
self-consistently determined by coupling to the electrons near the Fermi
level in $H_{\mathrm{eff}}$.

\emph{Effective theory. }One may redefine $\mathbf{n}_{i}$ as the new $%
\mathbf{\hat{z}}$-axis for the spin index of the electron spinor operator:
\begin{equation}
\hat{c}_{ia}=U_{i}\hat{a}_{ia}  \label{c}
\end{equation}%
with $\hat{a}_{ia}^{\dagger }=(\hat{a}_{ia,\uparrow }^{\dagger },\hat{a}%
_{ia,\downarrow }^{\dagger })$ by an SU(2) rotation%
\begin{equation}
U_{i}^{\dagger }\mathbf{n}_{i}\cdot \mathbf{\hat{\sigma}}U_{i}=\hat{\sigma}%
_{z}.  \label{U}
\end{equation}%
Then $H_{\mathrm{I}}$ simply reduces to%
\begin{equation}
H_{\mathrm{I}}=-\sum_{i,a}M_{a}(-1)^{i}S_{ia}^{z}  \label{Hi1}
\end{equation}%
while $H_{\mathrm{band}}$ becomes%
\begin{equation}
H_{\mathrm{band}}=-\sum_{ij,a,b}t_{ij}^{a,b}\hat{a}_{ia}^{\dagger }\left(
U_{i}^{\dagger }U_{j}\right) \hat{a}_{jb}  \label{hband-1}
\end{equation}%
We may further rewrite
\begin{equation}
H_{\mathrm{eff}}=H_{0}+H_{1}  \label{heff0}
\end{equation}%
where $H_{0}\equiv H_{\mathrm{I}}+H_{\mathrm{band}}[U_{i}^{\dagger }U_{j}=1]$
is simply an SDW mean-field Hamiltonian for the multibands, which can be
diagonalized as
\begin{eqnarray}
H_{0} &=&\sum_{\mathbf{k},a}\xi _{\mathbf{k}}^{a+}\left( \hat{\alpha}_{%
\mathbf{k}a}^{\dagger }\hat{\alpha}_{\mathbf{k}a}+\hat{\beta}_{\mathbf{k}%
a}^{\dagger }\hat{\beta}_{\mathbf{k}a}\right)   \nonumber \\
&&-\sum_{\mathbf{k},a}E_{\mathbf{k}}^{a}\left( \hat{\alpha}_{\mathbf{k}%
a}^{\dagger }\hat{\alpha}_{\mathbf{k}a}-\hat{\beta}_{\mathbf{k}a}^{\dagger }%
\hat{\beta}_{\mathbf{k}a}\right) +\mathrm{const.}  \label{H0}
\end{eqnarray}%
by a canonical transformation $\hat{a}_{\mathbf{k}a}=u_{\mathbf{k}}^{a}\hat{%
\alpha}_{\mathbf{k}a}-v_{\mathbf{k}}^{a}\hat{\sigma}_{z}\hat{\beta}_{\mathbf{%
k}a}$, $\hat{a}_{\mathbf{k+Q}a}=v_{\mathbf{k}}^{a}\hat{\sigma}_{z}\hat{\alpha%
}_{\mathbf{k}a}+u_{\mathbf{k}}^{a}\hat{\beta}_{\mathbf{k}a}$ with $u_{%
\mathbf{k}}^{a}=\left[ (1-\xi _{\mathbf{k}}^{a-}/E_{\mathbf{k}}^{a})/2\right]
^{1/2}$, $v_{\mathbf{k}}^{a}=\left[ (1+\xi _{\mathbf{k}}^{a-}/E_{\mathbf{k}%
}^{a})/2\right] ^{1/2}$. Here $\xi _{\mathbf{k}}^{a\pm }\equiv \left(
\varepsilon _{\mathbf{k}}^{a}\pm \varepsilon _{\mathbf{k+Q}}^{a}\right) /2$
and
\begin{equation}
E_{\mathbf{k}}^{a}=\sqrt{\left( \xi _{\mathbf{k}}^{a-}\right) ^{2}+\left(
M_{a}/2\right) ^{2}}  \label{spectrum E}
\end{equation}%
where $\varepsilon _{\mathbf{k}}^{a}$ denotes the bare spectrum determined
by (\ref{hband}) (setting the chemical potential $\mu =0$). Note that the
band label $a$ here can be different from the original orbital label in (\ref%
{hband}) because of the mixture of orbitals, and in obtaining (\ref{spectrum
E}), the same $M_{a}$ is assumed for the mixed orbitals. Here $\mathbf{k}$
is defined in the magnetic Brillouin zone (BZ)\ with the magnetic momentum $%
\mathbf{Q}=(\pi ,\pi )$, which coincides with the BZ of two irons per unit
cell (Fig. 2(b))$.$\

Now let us consider the term with $U_{i}^{\dagger }U_{j}\neq 1$:
\begin{equation}
H_{1}=-\sum_{ij,a,b}t_{ij}^{a,b}\hat{a}_{ia}^{\dagger }\left( U_{i}^{\dagger
}U_{j}-1\right) \hat{a}_{jb}  \label{H1}
\end{equation}%
We shall focus on the case in which the $\alpha $-bands are all filled up by
the electrons, and the Fermi energy is located in some of $\beta $-bands,
which corresponds to both the undoped and electron-doped (or slightly
hole-doped) situations as illustrated by Fig. 1. After integrating out the $%
\alpha $-bands and by assuming $\Delta _{\alpha }\mathbf{n}_{i}$ is small, $%
H_{1}$ may be simplified\cite{weng90} (at large $M_{a}$) to
\begin{eqnarray}
H_{1} &\simeq &\frac{1}{2}\sum_{\mathbf{k},\mathbf{q,}\sigma }\nabla
\varepsilon _{\mathbf{k}}\cdot \left( \mathbf{D}_{\mathbf{q}}\right)
_{\sigma ,-\sigma }\left( -\theta _{\mathbf{k}+\mathbf{q}}+\sigma \bar{\theta}%
_{\mathbf{k}+\mathbf{q}}\right) \beta _{\mathbf{k+q,}\sigma }^{\dagger
}\beta _{\mathbf{k,-}\sigma }  \nonumber \\
&&+\text{ }\mathrm{H.c.}+\frac{J_{\mathrm{eff}}}{8}\sum_{i}\left( \Delta
\mathbf{n}_{i}\right) ^{2}  \label{h1-2}
\end{eqnarray}%
where $\mathbf{D}_{\mathbf{q}}$ is a Fourier transformation of

\begin{equation}
\mathbf{D}_{i}=\frac{1}{2}(\mathbf{\Delta n}_{i}\times \mathbf{n}_{i})\cdot
U_{i}^{\dagger }\mathbf{\hat{\sigma}}U_{i}  \label{D}
\end{equation}%
and $\theta _{\mathbf{k}+\mathbf{q}}$($\bar{\theta}_{\mathbf{k}+\mathbf{q}}$%
) is the step function restricting $\mathbf{k}+\mathbf{q}$ within (outside)
the magnetic BZ. Note that for simplicity we have omitted the band indices
in (\ref{h1-2}), where the spiral field $\mathbf{D}_{\mathbf{q}}$ will
couple to the electrons from all $\beta $-bands near the Fermi level. It
shows that the twist of $\mathbf{n}_{i}$ is kinetic energy driven. Such a
spiral twist is balanced by the \textquotedblleft
superexchange\textquotedblright\ term in $H_{1}$, where $J_{\mathrm{eff}%
}\simeq 1/N\sum_{\mathbf{k}a}(1-n_{\mathbf{k}a}^{\beta }$)$\left( \xi _{%
\mathbf{k}}^{a-}\right) ^{2}/E_{\mathbf{k}}^{a}$ provides the spin stiffness
against the twist of the SDW order and is mainly contributed by the filled $%
\alpha $-bands and is reduced with the increasing filling $n_{\mathbf{k}%
}^{\beta }$ in the $\beta $-bands.

Some technical remarks with regarding the derivation of (\ref{h1-2}) are in
order. Strictly speaking, the dynamic field $\mathbf{n}_{i}$ should be
introduced in a path-integral formalism,\cite{weng90,schulz} where the
temporal term $U_{i}^{\dagger }\mathbf{\partial }_{t}U_{i}$ will also enter
the Lagrangian. As shown previously for the one-band Hubbard model,\cite%
{weng90,weng91} the terms like $U_{i}^{\dagger }\mathbf{\partial }_{t}U_{i}$
will play an important role in determining the dynamics of $\mathbf{n}_{i}$
at half-filling where an SDW long-range order is present. In this case, by
expanding $\mathbf{n}_{i}$ around $\mathbf{\hat{z}}$, $\mathbf{D}_{\mathbf{q}%
}\simeq i\mathbf{q}(\mathbf{n}_{\mathbf{q}}\times \mathbf{\hat{z}})\cdot
\mathbf{\hat{\sigma}}$, the propagator of $\mathbf{D}_{\mathbf{q}}$ reduces
to $\mathcal{D}^{s}(\mathbf{q},t)\sim \mathbf{qq}\left\langle \mathbf{n}_{%
\mathbf{q}}^{\perp }(t)\cdot \mathbf{n}_{\mathbf{q}}^{\perp
}(0)\right\rangle $ which is proportional to the spin-wave propagator with
the coupling to doped particles vanishing as $\mathbf{q}\rightarrow 0$.
However, once a finite density of the holes or electrons are doped into the
system, the dynamic spiral fluctuation of $\mathbf{D}_{q}$ which couples to
the doped charge carriers in (\ref{h1-2}) will become dominant over the
other fluctuation terms like $U_{i}^{\dagger }\mathbf{\partial }_{t}U_{i}$,
leading to intrinsic instabilities\cite{weng91} of the system (see below).

The propagator $\mathcal{D}(\mathbf{q},t)=-i\left\langle T_{t}\mathrm{Tr}%
\left[ \mathbf{D}_{\mathbf{q}}(t)\mathbf{D}_{-\mathbf{q}}(0)\right]
\right\rangle $ can be obtained by integrating out the $\beta $-band%
\begin{equation}
\mathcal{D}(\mathbf{q},\omega )\simeq -1/\left[ \mathbf{\Pi }_{\beta }%
\mathbf{(q},\omega )+J_{\mathrm{eff}}\right]  \label{propagator}
\end{equation}%
where $\mathbf{\Pi }_{\beta }\mathbf{(q},\omega )$ is the \textquotedblleft
bubble\textquotedblright\ diagram contribution of the $\beta $-band
electrons, with the \textquotedblleft bare\textquotedblright\ $\mathbf{\Pi }%
_{\beta }^{0}\mathbf{(0},0)\sim -v_{F}^{2}N_{F}\sim -t$ where $v_{F}$ and $%
N_{F}$ are the Fermi velocity and density of states based on $H_{0}$, which
is expected to dominate over $J_{\mathrm{eff}}$ on general grounds,\cite%
{weng91} reflecting the fact that the hopping energy is always dominant over
the superexchange energy locally, leading to the instability of the AF
long-range order.

\emph{\textquotedblleft Stripe\textquotedblright\ instability in the undoped
case. }According to the band structure calculations,\cite{cao,lu} in the
undoped case, there are hole-pockets around the $\Gamma $ point and electron
pockets around the $M$ point around the Fermi level, which are connected by
a particular nesting momentum $\mathbf{Q}_{s}$, and can be described by $%
H_{0}$ with the proper choice of $\varepsilon _{\mathbf{k}}^{a}$. Then, to
take advantage of the enhanced response function at $\mathbf{Q}_{s}$ in $%
H_{0}$, the interaction term $H_{1}$ will naturally induce

\begin{equation}
\nabla \varepsilon _{\mathbf{k}}\cdot \left\langle \mathbf{D}_{\mathbf{Q}%
_{s}}\right\rangle \neq 0  \label{stripe order}
\end{equation}%
and result in a mean-field $\left\langle \beta _{\mathbf{k+Q}_{s},\sigma
}^{\dagger }\beta _{\mathbf{k,-}\sigma }\right\rangle \neq 0$ with a $\sqrt{2%
}\times \sqrt{2}$ folding of the BZ as shown in Fig. 2(b). Consequently a
gap will be opened up at the Fermi energy to stabilize $\mathcal{D}(\mathbf{q%
},\omega )$ in (\ref{propagator}). In contrast to the unstable SDW with the
AF momentum $\mathbf{Q}$, the true magnetic ordering (\ref{stripe order}) is
realized with an additional spiral twist at momentum $\mathbf{Q}_{s}$,
which, known as a stripe-type SDW, was first predicted in Ref. \cite{zfang}
and has been recently confirmed experimentally by neutron scattering.\cite%
{cruz,McGuire}

\emph{Superconducting instability. } With introducing a small amount of
doped electrons, the Fermi energy will move up such that the small hole
pockets around the $\Gamma $ point get filled up and the Fermi surface
nesting disappears. Then the static \textquotedblleft
stripe\textquotedblright\ SDW order (\ref{stripe order}) vanishes, and $%
\mathbf{D}_{\mathbf{q}}$ will become unstable again.

Rewriting
\begin{equation}
U_{i}^{\dagger }U_{j}=U_{i}^{\dagger }U_{i_{1}}U_{i_{1}}^{\dagger
}U_{i_{2}}U_{i_{2}}^{\dagger }...U_{i_{N_{ij}-1}}^{\dagger }U_{j}  \label{U1}
\end{equation}%
with $\{i_{s}\}=i_{0},i_{1,}...,i_{N_{ij}}$ a sequence of lattice sites
connecting $i\equiv i_{0}$ and $j\equiv i_{N_{ij}}$ and using $%
U_{i_{s}}^{\dagger }U_{i_{s+1}}\simeq 1+i\mathbf{D}_{i_{s}}\cdot \mathbf{%
\hat{\eta}}_{i_{s+1},i_{s}}\simeq e^{i\mathbf{D}_{i_{s}}\cdot \mathbf{\hat{%
\eta}}_{i_{s}+1,i_{s}}}$ ($\mathbf{\hat{\eta}}_{i_{s}+1,i_{s}}=\mathbf{r}%
_{i_{s}+1}-\mathbf{r}_{i_{s}}$) (neglecting the phase associated with the
solid angle spanned by $\mathbf{n}_{i_{s}}$,$\mathbf{n}_{i_{s+1}}$, $\mathbf{%
\hat{z})}$, one finds%
\begin{eqnarray}
\left\langle U_{i}^{\dagger }U_{j}\right\rangle &\sim &\left\langle \exp
i\int\nolimits_{i}^{j}d\mathbf{r}\cdot \mathbf{D}\right\rangle  \nonumber \\
&\simeq &\exp \left[ -\frac{1}{2}\int \int\nolimits_{i}^{j}d\mathbf{r}\cdot
\left\langle \mathbf{DD}\right\rangle \cdot d\mathbf{r}^{\prime }\right]
\nonumber \\
&\simeq &\exp \left[ -\frac{\left\vert i-j\right\vert ^{2}}{2\xi _{s}^{2}}%
\right]  \label{U2}
\end{eqnarray}%
where the spin correlation length $\xi _{s}\sim 1/\left[ -\int d\omega Im%
\mathcal{D}(0,\omega )\right] ^{1/2}$. \ Unless $Im\mathcal{D}(\mathbf{q}%
,\omega )\rightarrow 0$ at $\mathbf{q}\rightarrow 0$ as in the spin wave
case, the spiral fluctuations in (\ref{propagator}) will generally lead to a
spin disordered state with a finite $\xi _{s}$, but its precise nature has
to be self-consistently determined in view of the divergent $\mathcal{D}(%
\mathbf{q},\omega )$ in the small $\mathbf{q}$ and $\omega $ regime.

The dynamic spiral fluctuations of the magnetization directions are kinetic
energy driven by the electrons near the Fermi level. However, this will also
make the electrons lose their long-wavelength coherence. Indeed, the
single-particle propagator for the electrons to leading order of
approximation can be written as%
\begin{equation}
G^{e}(i,j;t)\simeq \left\langle U_{i}^{\dagger }(t)U_{j}(0)\right\rangle
G^{a}  \label{G}
\end{equation}%
where $G^{a}$ is the propagator for the $a$-particles whose leading term is
coherent governed by the mean-field Hamiltonian $H_{0}$ in (\ref{H0}). Thus
the dynamic spiral fluctuation will quickly damp the coherent motion of the
quasiparticle beyond the spin correlation length $\xi _{s}$ via $%
\left\langle U_{i}^{\dagger }(t)U_{j}(0)\right\rangle $.

By contrast, after averaging over the $\mathbf{n}_{i}$-field, the electron
singlet pair operator $\hat{\Delta}_{ij}^{SC}\equiv \sum_{\sigma }\sigma
c_{i\sigma }^{\dagger }c_{j-\sigma }^{\dagger }$ can be expressed by%
\begin{eqnarray}
\Delta _{ij}^{SC} &\simeq &\left\langle U_{i}^{\dagger }U_{j}\right\rangle
\sum_{\sigma }\sigma a_{i\sigma }^{\dagger }a_{j-\sigma }^{\dagger }
\nonumber \\
&\approx &\sum_{\sigma }\sigma a_{i\sigma }^{\dagger }a_{j-\sigma }^{\dagger
}  \label{Dsc}
\end{eqnarray}%
at $\left\vert i-j\right\vert \ll \xi _{s}$. It clearly shows that in such a
spin disordered state, the singlet pairs of electrons can still propagate
coherently. The above contrast between the single electron and singlet pairs
of electrons indicates an instability of the system towards
superconductivity with $\sum_{\sigma }\sigma \left\langle a_{i\sigma
}^{\dagger }a_{j-\sigma }^{\dagger }\right\rangle \neq 0$. The latter can be
indeed realized by exchanging the dynamic spiral fluctuations$,$ $\mathcal{D}%
(\mathbf{q},\omega )$, between the $\beta $-electrons based on (\ref{h1-2}),
which in turn stabilizes $\mathcal{D}(\mathbf{q},\omega )$ through a
renormalized $\mathbf{\Pi }_{\beta }\mathbf{(q},\omega )$ as the $\beta $%
-electrons form Cooper pairs. It is noted that the stabilized $\mathrm{Re}%
\mathcal{D}(\mathbf{q},\omega )<0$ at small $\mathbf{q}$ and $\omega $ will
lead to an attractive (repulsive) interaction between $\beta $-electrons if $%
\mathbf{k}+\mathbf{q}$ is within (outside) the magnetic BZ according to (\ref%
{h1-2}). It means that the $\beta $-electrons will form the dominantly d$%
_{x^{2}-y^{2}}$-wave pairing at four $M$ points in Fig. 2(b), similar to the
case in the cuprates. Physically this kinetic-energy-driven pairing can be
understood as that two electrons sitting at, say, NN sites, gain enhanced
hopping energies by sharing the spiral twist of $\mathbf{n}_{i}$'s between
the two sites.

\emph{Discussions. }It is interesting to point out that $T_{c}$ is upper
bounded in the BCS theory because, whereas the softening phonon can enhance
the attractive force, it also leads to the structural instability of the
solids. Here the superconductivity is caused by strong dynamic spiral
fluctuations which are in a \textquotedblleft melting\textquotedblright\ SDW
regime. $T_{c}$ is expected to fall when such dynamic spiral fluctuations
get reduced at higher electron doping concentration (the local magnetization
$M_{a}$ should be eventually destroyed when the Fermi level reaches beyond
the highest $\beta $-band shown in Fig. 1, presumably the one with the
dominant $d_{x^{2}-y^{2}}$ characters according to the LDA calculation\cite%
{lu}).

Although the d-electrons in Fe-based compounds are believed quite itinerant,
our proposal suggests that the underlying physics is still far from the
conventional itinerant magnetic metals in the following sense. Due to the
presence of a large \emph{local} SDW mean-field $M_{a}$, each band near the
Fermi level are still split into the lower and upper Hubbard bands, with the
lower one filled by the electrons in both undoped and electron-doped cases,
which are responsible for the origin of large $M_{a}$'s. In contrast to the
single-band Hubbard model relevant to the cuprate superconductors,\cite%
{anderson} however, here some of the upper Hubbard bands are already at the
Fermi level even in the undoped case. A search for the depleted density of
states below the Fermi level by, say, photoemission,\cite{feng} may provide
useful information concerning the correctness of the present model. The
presence of a sizable local moment, which is distinguished from the
itinerant approach, may be also investigated via various magnetic
measurement above $T_{c}$, including the magnetic susceptibility.\cite%
{chenxy2}

Finally we caution that in the present model the effect of the unit cell
doubling with two irons per cell due to the crystal field reason (As ions
are displaced above and below the Fe plane alternatively) has not been
considered, which may drastically affect our results if it becomes
sufficiently strong because in that case the $\alpha $- and $\beta $-band
splitting due to the local SDW may no longer be complete and $\alpha $-bands
can become partially filled.

In conclusion, we have proposed a simple effective model to describe the
low-energy physics in FeAs superconductors. This model is based on an
assumption that there exists a robust local SDW order, but its magnetization
direction is intrinsically unstable against forming a true long range order
due to the competition with the kinetic energy of the charge carriers near
the Fermi energy. In the undoped case, a weak SDW order of stripe type is
formed due to the peculiar nesting structure of the Fermi surfaces. In the
doped case, the dynamic melting of the SDW order will result in the d-wave
superconducting pairing of the doped electrons near the Fermi energy.

\bigskip

\textbf{Acknowledgments:} Stimulating discussions with Z.Y. Lu, Z. Fang, K.
Wu, J.W. Mei, T. Li are acknowledged. We also thank N.L. Wang, H.H. Wen, and
especially X.H. Chen for generously sharing their experimental results
before publication. This work is supported by NSFC grant no. 10688401.

\vspace*{2mm}

\newpage

\bigskip

\bigskip


\begin{thebibliography}{99}
\bibitem{yoichi} Y. Kamihara, et al., \emph{J. Am. Chem. Sco.} \textbf{128},
10012 (2006); Y. Kamihara, et al., \emph{J. Am. Chem. Sco.} \textbf{130},
3296 (2008).

\bibitem{wen} H.-H. Wen, et al., Europhys. Lett. \textbf{82}, 17009 (2008).

\bibitem{chenxh} X. H. Chen, et al., arXiv:0803.3603 (2008).

\bibitem{nlwang} G. F. Chen, et al., arXiv:0803.3790 (2008).

\bibitem{ren} Z. A. Ren, et al., arXiv:0803.4283 (2008); arXiv:0804.2053
(2008);

\bibitem{zfang} J. Dong, et al., arXiv:0803.3426 (2008).

\bibitem{cruz} C. Cruz, et al., arXiv:0804.0795(2008).

\bibitem{McGuire} M. A. McGuire, et al., arXiv:0804.0796 (2008).

\bibitem{chenxh1} R. H. Liu, et al., arXiv:0804.2105 (2008).

\bibitem{lebegue} S. Leb\`{e}gue, Phys. Rev. B \textbf{75}, 035110 (2008).

\bibitem{singh} D. Singh and M. Du, arXiv:0803.0429 (2008); I. Mazin, et
al., arXiv:0803.2740 (2008).

\bibitem{kotlia} K. Haule, et al., arXiv:0803.1279 (2008).

\bibitem{xu} G. Xu, et al., arXiv:0803.1282 (2008).

\bibitem{cao} C. Cao, et al., arXiv:0803.3236(2008).

\bibitem{lu} F. Ma and Z. Y. Lu arXiv:0803.3286(2008).

\bibitem{kuroki} K. Kuroki, et al., arXiv:0803.3325 (2008).

\bibitem{dai} X. Dai, et al., arXiv:0803.3982 (2008).

\bibitem{zdwang} Q. Han, et al., Europhys. Lett. \textbf{82}, 37007 (2008).

\bibitem{qi} S. Raghu, et al., arXiv:0804.1113 (2008).

\bibitem{lee} P. A. Lee and X. G. Wen, arXiv:0804.1113 (2008).

\bibitem{li} T. Li, arXiv:0803.1279 (2008); G. Baskaran, arXiv:0803.1341
(2008); Q. M. Si and E. Abrahams, arXiv:0804.2480 (2008).

\bibitem{anderson} P. W. Anderson, Science \textbf{235}, 1196 (1987).

\bibitem{weng90} Z. Y. Weng, et al., Phys. Rev. B\textbf{43},3790 (1990).

\bibitem{schulz} H. J. Schulz, Phys. Rev. Lett. \textbf{65}, 2462 (1990).

\bibitem{weng91} Z. Y. Weng, Phys. Rev. lett. \textbf{66}, 2156 (1991).

\bibitem{feng} H.W. Ou, et al., arXiv:0803.4328 (2008).

\bibitem{chenxy2} X. H. Chen, unpublished.
\end{thebibliography}
\end{document}